\documentclass[page-classic]{epl2}
\usepackage{latexsym}
\usepackage{amstext,amsfonts,amssymb}
\usepackage{epsfig}

\title{Disorder Induced Limited Path Percolation}
\shorttitle{Disorder Induced Limited Path Percolation} 

\author{Eduardo L\'opez\inst{1,2} \and Lidia A. Braunstein\inst{3,4}}
\shortauthor{E. L\'opez \etal}

\institute{                    
  \inst{1} CABDyN Complexity Centre, Sa\"{\i}d Business School, University of Oxford, Park End Street,
OX1 1HP, United Kingdom\\
  \inst{2} Physics Department, Clarendon Laboratory, University of Oxford, Parks Road, Oxford OX1 3PU, 
United Kingdom\\
  \inst{3} Instituto de Investigaciones F\'{\i}sicas de Mar del Plata (IFIMAR), Departamento de F\'{\i}sica, 
Facultad de Ciencias Exactas y Naturales, Universidad Nacional de Mar del Plata-CONICET, Funes 3350, (7600) 
Mar del Plata, Argentina\\
  \inst{4} Center for polymer studies, Boston University, Boston, MA 02215, USA
}

\abstract{
We introduce a model of percolation induced by disorder, 
where an initially homogeneous network with
links of equal weight is disordered by the introduction of heterogeneous weights for the links.
We consider a pair of nodes $i$ and $j$ to be mutually reachable when the ratio $\alpha_{ij}$ of length
of the optimal path between them before and after the introduction of
disorder does not increase beyond a tolerance ratio $\tau$. 
These conditions reflect practical limitations of reachability better than the usual percolation model,
which entirely disregards path length when defining connectivity and, therefore, communication.
We find that this model leads to a first order phase transition in both 2-dimensional lattices and
in Erd\H{o}s-R\'enyi networks, and in the case of the latter, the size of the discontinuity 
implies that the transition is effectively catastrophic, with almost all system pairs undergoing
the change from reachable to unreachable.
Using the theory of optimal path lengths under disorder, we are able to predict the percolation threshold.
For real networks subject to changes while in operation, this model should perform better in predicting
functional limits than current percolation models.}

\begin{document}

\maketitle

Percolation theory is the most well-established approaches to
study system connectivity, and how this connectivity becomes compromised 
with local system failure~\cite{Percolation}. Systems usually refer 
to connected structures such as lattices or random networks~\cite{rev-Albert}, 
and failures to the removal of nodes or links.
Percolation predicts failure thresholds and size of the
connected parts of the network after those failures. Practical domains of applicationes include  
epidemiology~\cite{Grassberger, Pastor-Satorras, Cohen}, communication networks
like the Internet~\cite{Barabasi-attacks}, and propagation of
information in social networks~\cite{Onnela}.

Recently~\cite{Lopez-LPP}, it was pointed out that there are many contexts in which structural
connectivity may not be sufficient for a network to maintain its functionality when faced with failures.
Network nodes may need to be more than connected; they may need to be reachable from one another, which is 
practical definition.
In regular percolation, node $j$ is reachable from node $i$ if there is a path of consecutive
links from $i$ to $j$, i.e., if the nodes are structurally connected after any failures.
However, in real networks, two nodes initially reachable from one another through a structural path
can become unreachable after link failures even if there is a new 
structural path connecting them, but which is in some way unsatisfactory. 
This reasoning lead to the introduction of Limited Path Percolation (LPP), in which 
reachability between nodes $i$ and $j$ after the onset of failures is a relative statement: if
the path length between nodes before ($\ell_{ij}$) and after ($\ell^{'}_{ij}$) failures is
such that $\ell^{'}_{ij}/\ell_{ij}\leq\tau$, then node $j$ is reachable from $i$ (and vice versa), where $\tau$
is an externally imposed tolerance that reflects properties of the functional aspect of the network 
(regular percolation corresponds to $\tau\rightarrow\infty$).
For a large class of networks, Ref.~\cite{Lopez-LPP} shows that a new phase transition appears, 
where networks are now more fragil due to the additional length constraint.

The approach introduced in Ref.~\cite{Lopez-LPP} was restricted to the 
removal (absolute failure) of some links, while other links remained unperturbed.
However, in many real-world examples, links usually do not fail entirely,
but can instead become more difficult to use such as in road networks affected by weather. 
One can then imagine a network to begin with no performance degradation
(equally useful links forming an ``ordered'' network), 
but progressively develop some form of disorder 
in which each link becomes costly to use (with cost or weight $w$). 
In other words, the network acquires a distribution of link weights 
$P_a(w)$ where $a$ is the disorder parameter~\cite{no-time-depen}. 
Once disorder has set in, new ``optimal'' 
paths (paths of least total weight) between each pair of nodes must be found, and a new {\it disorder induced
Limited Path Percolation} can be defined on the basis of the path length change between the original
unweighted network and the subsequent weighted network,
reflecting a scenario more akin to many real networks.

In this article, we study disorder induced LPP, and find that 
the model leads to a first order phase transition for both lattices and Erd\H{o}s-R\'enyi networks,
indicating a catastrophic failure of the system when the disorder exceeds the tolerance limit of the network.
For Erd\H{o}s-R\'enyi networks, the failure appears to be catastrophic, as our numerical
results indicate that the phase transition occurs between an almost fully connected network to
a network of fractional size zero in the thermodynamic limit.
Through the use of theoretical results that explain the behavior of optimal paths,
we are able to predict the location of the LPP transition in both network structures
as a function of the disorder, and structural properties of the system.
Our model also displays universal features with respect to the disorder, as
demostrated by our numerical results.
\section{Model and Methods}
\subsection{Formulation of Limited Path Percolation induced by disorder}
To be concrete, LPP in Ref.~\cite{Lopez-LPP}, where link failure corresponds to link removal,
is formulated in the following way: if a pair of 
nodes $i$ and $j$ is connected through a shortest path of length $\ell_{ij}$ (number of links) 
at $p=1$ (no links removed), and of length $\ell^{'}_{ij}(p)$ at $p<1$ (fraction $1-p$ of links removed) , 
$i$ and $j$ are considered reachable if
$\ell^{'}_{ij}(p)\leq \tau\ell_{ij}$, where $\tau$ is the tolerance 
factor which lies on the range between 1 and $\infty$~\cite{Lopez-LPP}. 
The number of reachable pairs, $S$, is measured through 
\begin{equation}
S=\sum_{i\neq o} \theta(\tau\ell_{oi}-\ell^{'}_{oi})
\label{S}
\end{equation}
where $o$ is an ``origin'' node chosen to minimize its impact on $S$ (see below), and 
$\theta$ is the Heaveside step function $\theta(x)=1$ is $x\geq 0$, and 0 otherwise.
$S$ depends on both $p$ and $\tau$, and the
LPP phase transition occurs for the combination of these parameters at the threshold of the relation $S\sim N$. 
For fixed $\tau$, there is a threshold $p=\tilde{p}_c(\tau)$
at which $S\sim N$ is achieved. Alternatively, given $p\geq p_c$~\cite{fnpgtrpc}, there is a
$\tau=\tau_c(p)$ such that $S\sim N$. When $\tau\rightarrow\infty$, $\tilde{p}_c(\tau\rightarrow\infty)=p_c$.

For disorder induced LPP, the set up is similar, but instead of considering 
each link to be kept with probability $p$, we 
change each link weight from 1 to $w$ drawn from a random uncorrelated distribution $P_a(w)$
where $a$ is the disorder parameter (defined below)~\cite{reg-LPP}. 
The path lengths change from $\ell_{ij}$ to $\ell^{'}_{ij}(a)$ where now the later
corresponds to the length of the optimal path between $i$ and $j$. 
Reachability is then defined as in Ref.~\cite{Lopez-LPP}: $i$ and $j$ are considered
reachable if $\ell^{'}_{ij}(a)\leq \tau\ell_{ij}$. The number of node pairs $S$ that remain
reachable is a function of $a$, $\tau$, and $N$. We search for the LPP 
threshold by imposing $S\sim N$. If a threshold exists
for a fixed tolerance $\tau$, there is a critical disorder $a=a_c(\tau)$ or,
vice versa, a critical tolerance $\tau=\tau_c(a)$ that depends on the disorder parameter $a$.
Another way to phrase this is to say that $\tau$ and $a$ are control parameters in disorder
induced LPP, as $\tau$ and $p$ are in regular LPP.

To consider a tolerance to the path length increase without considering an associated tolerance to the 
path weight increase may at first seem arbitrary, but in fact 
is well justified in that the overall weight of a path is asymptotically proportional to its length
under the conditions of disorder we study here. Hence, choosing path length tolerance does
not affect the qualitative nature of our results, and one choice of tolerance can be mapped
onto the other. In concrete terms, if we imagined disorder corresponding to something like
time of travel through a link, then total average travel time scales linearly with the trip distance.

The general algorithm used to measure disorder induced LPP is the following. 
First, we select the ensemble of networks $\mathcal{G}$ of interest. In this article
we focus on 2-dimensional square lattices and Erd\H{o}s-R\'{e}nyi (ER) graphs~\cite{ER}. 
Square lattices have no randomness, of course, but can be formally viewed as an ensemble
with one single realization.
For each network realization $G\in\mathcal{G}$, in which all links have weight 1,
we determine the path length $\ell_{oi}$ between nodes $o$ and $i$ for all $i\neq o$. 
On the same network realization $G$ (the same nodes and links), 
disorder is introduced by changing the weight of each link from 1 to $w_{ij}$. Subsequently, 
the optimal paths between $o$ and all other nodes $i$ of $G$ are determined, and their lengths 
$\ell^{'}_{oi}$ recorded. $S$ is calculated by using Eq.~(\ref{S}). 
To determine path lengths, we use the Dijkstra algorithm~\cite{Cormen}. 
\subsection{Disorder}
We consider disorder distributions $P_a(w)$ characterized by a single
disorder parameter which, for convenience, we label as $a$. Generally, we would expect
that disorder induced LPP changes as a function of the form of $P_a(w)$.
However, recent work~\cite{Chen} shows
that large classes of disorder distributions are essentially equivalent in the optimal path 
problem, provided a certain characteristic length scale $\xi(a)$ associated with $P_a(w)$ is conserved
(see below).
In practical terms, this means that disorder distributions of different functional forms but with
equal $\xi(a)$ lead to optimal path distributions that scale in the same way~\cite{Chen}.

This result allows us to choose
a distribution that is convenient and well understood. Thus, we use
\begin{equation} 
P_a(w)=(aw)^{-1},\qquad [w\in [1,e^a]]
\label{Pw}
\end{equation}
for which a large amount of research has been conducted in the context of the optimal path
problem~\cite{Cieplak,Porto,Braunstein,Wu,Ambegaokar}. 
Determining $\xi$ for a given distribution is addressed in~\cite{Chen}, and we 
return to this in the discussion of results. 
\section{Results and Discussion}
To characterize LPP effectively, we first measure $\Theta(S|\tau,a,N)$, the distribution of sizes of 
the cluster containing $o$, for networks of size $N$ with tolerance $\tau$
over disorder and network realizations. 
Previously~\cite{Lopez-LPP}, the LPP transition was found by calculating 
$\langle S\rangle=\sum_{S} S\Theta(S)$ and determining the parameter values at the threshold of
$\langle S\rangle\sim N$. Here we develop a more systematic approach, in 
which we look at the entire distribution $\Theta(S|\tau,a,N)$
in order to explore whether disorder induced LPP exhibits a phase transition, and if
it does, what is the order.
To study the thermodynamic limit, we
find it useful to analyze the fractional mass $\sigma\equiv S/N$, and hence 
$\Theta(\sigma|\tau,a,N)$ (or $\Theta(\sigma|\tau,a,L)$ for lattices).

\begin{figure}
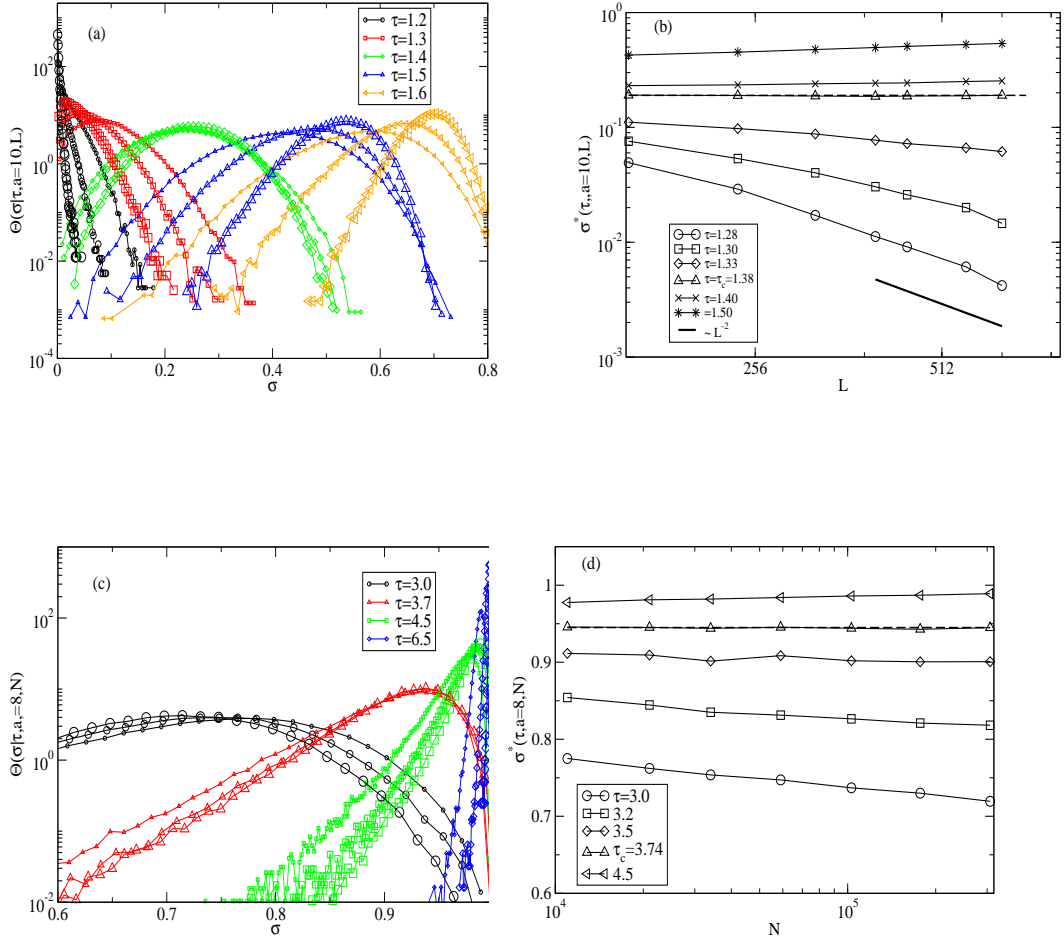

\epsfig{file=fig1a.eps,width=6.5cm,height=5.2cm}\vspace{1cm}
\epsfig{file=fig1b.eps,width=6.5cm,height=5.2cm}\vspace{1cm}
\epsfig{file=fig1c.eps,width=6.5cm,height=5.2cm}\vspace{1cm}
\epsfig{file=fig1d.eps,width=6.5cm,height=5.2cm}
\caption{(a)$\Theta(\sigma|\tau,a=10,L)$ vs. $\sigma$ for the square lattice 
for various $\tau$ (in legend) and $L=160,320,640$ (increasing symbol size corresponds to increasing system size).
Note that for $\tau=1.4$ the peaks of the distributions $\sigma^*$ remain almost 
in the same location with increasing $N$ indicating that $\tau_c^{\rm(Latt)}$ is near this value. 
For $\tau<\tau_c^{\rm(Latt)}$, $\sigma^*$ decreases with $N$, and
for $\tau>\tau_c^{\rm(Latt)}$, $\sigma^*$ increases with $N$.
Further detailed simulations reveal a better estimate of $\tau_c^{\rm(Latt)}$ is 1.38 for $a=10$.
From $\Theta(\sigma|\tau,a,L)$, we determine $\sigma^*$ by cubic regression of the top points
of the peak of $\Theta$.
(b) $\sigma^*(\tau,a=10,L)$ vs. $L$ for square lattice, for several $\tau$ and $L$. With 
$\tau=\tau_c^{\rm(Latt)}=1.38$
the curve stays constant indicating the phase transition; for $\tau$ approaching 1 (path lengths cannot increase much), 
$\sigma^*$ scales as $L^{-2}$ (solid line), indicating that the largest reachable component is logarithmic in size.
(c) $\Theta(\sigma|\tau,a=8,N)$ vs. $\sigma$ for Erd\H{o}s-R\'enyi networks of $\langle k\rangle=3$,
$N=11000, 59000, 307000$ (increasing symbol size corresponds to increasing system size), 
and various $\tau$. In this case, for $\tau\approx 3.7$,
$\sigma^*$ stays roughly constant with increasing $N$, indicating $\tau_c^{\rm(ER)}$ is close to this value.
A more detailed analysis reveals that $\tau_c^{\rm(ER)}=3.74$ for $a=8$ and $\langle k\rangle=3$.
(d) $\sigma^*(\tau,a=8,N)$ for ER networks, with $\tau$ values indicated in the legend, and
$N$ ranging from 11000 to 307000. It is clear that $\sigma^*$ increases with $N$ 
for $\tau>\tau_c^{\rm(ER)}$ and decreases for $\tau<\tau_c^{\rm(ER)}$.}
\label{ThetaS_vs_N}
\end{figure}

On a square lattice of equal sides $L$ and $N=(L+1)^2$ nodes, we measure
$\Theta(\sigma|\tau,a,L)$ from node $o$ located at
the center of the lattice $(x_o=[[(L+1)/2]],y_o=[[(L+1)/2]])$ with $[[.]]$ indicating the next lowest integer of
the argument.
In Fig.~\ref{ThetaS_vs_N}(a), we show $\Theta(\sigma|\tau,a=10,L)$
with several $L$ and $\tau$. The first interesting feature is the narrow shape of the distribution,
indicating the presence of a characteristic mass $S$ for given $\tau$ and $L$. This suggests focusing on 
the most probable value of $\sigma$, labeled $\sigma^*(\tau,a,N)$, i.e. 
$\Theta(\sigma^*|\tau,a,N)>\Theta(\sigma|\tau,a,N)$ for all $\sigma\neq\sigma^*$.
Also, we observe that for small $\tau$, as the system size increases, $\sigma^*$
systematically decreases, but in contrast, for large $\tau$, $\sigma^*$ increases.
Between these two cases, we find a $\tau$ for which $\sigma^*(\tau)$ remains virtually constant.
For such $\tau$, labeled $\tau_c^{(\rm Latt)}$, $S\sim N$ since $\sigma^*(\tau_c)=const.$, 
signaling the appearance of a phase transition. 
From Fig.~\ref{ThetaS_vs_N}(a) we observe that $\sigma^*(\tau_c^{\rm (Latt)})$ is close to
0.20, suggesting a first order transition.

A systematic study of $\sigma^*(\tau,a,N)$ can be carried out with the purpose of understanding in
more detail the phase diagram of the model. In Fig.~\ref{ThetaS_vs_N}(b) we present $\sigma^*$ for $a=10$
and a range of $L$ and $\tau$. The value of $\sigma^*$ was estimated
from $\Theta(\sigma|\tau,a,L)$ by finding an appropriate cubic fit for the peak of the distribution
and calculating the location of its maximum. Three main regimes can be observed for $\tau$
above, close to, and below $\tau_c^{(\rm Latt)}$. 
The determination of $\tau_c^{(\rm Latt)}$ as a function of $a$ from simulations is done by inspection 
and requires exploring a range of $\tau$ with considerable precision ($\delta\tau\sim10^{-2}$ or
even $10^{-3}$, becoming more sensitive for large $a$) around a certain region, in plots such as Fig.~\ref{ThetaS_vs_N}(b).
For small $\tau$ (say, close to 1 and well below $\tau_c^{(\rm Latt)}$), it is interesting to see that as the
system size increases, $\sigma^*\sim L^{-2}$ indicating that the fraction of the system that is reachable 
is smaller than any power of $N$. The situation for $\tau\lesssim\tau_c^{(\rm Latt)}$ is not as clear:
in LPP due to link removal~\cite{Lopez-LPP} there is a regime of power law sizes of $S$, whereas here
such regime is not evident for lattices but seems to be present for ER networks (see below);
our current theory does not shed light on the matter.
For $\tau>\tau_c^{(\rm Latt)}$, we find a progressive increase of $\sigma^*$ with respect to $L$, with
a saturation value that depends on $\tau$; $\sigma^*$ gradually approaches 1 
as $\tau\rightarrow\infty$.

To measure $\Theta(\sigma|\tau,a,N)$ in ER networks, we sample over network realizations of $G\in\mathcal{G}$, 
and for each $G$ choose an origin $o$ at random. 
In Fig.~\ref{ThetaS_vs_N}(c) we present the relevant simulation results.
The qualitative features of $\Theta(\sigma|\tau,a,N)$ for lattices are also present in
ER networks, including the existence of a critical $\tau$, labeled $\tau_c^{(\rm ER)}$.
In contrast to lattices, $\sigma^*(\tau_c^{(\rm ER)})$ is close to 1, indicating a very
dramatic LPP transition, in which a slight change of $\tau$ around $\tau_c^{(\rm ER)}$ leads to a 
transition between almost entirely reachable to entirely unreachable global network states.
We also observe that for $\tau<\tau_c^{(\rm ER)}$, there seems to be a power-law decaying relation between
$\sigma^*$ and $N$, consistent with a fractal size object below the threshold, with the decay
exponent $\tau$-dependent. 

To analyze the problem further,
we define the quantity $\alpha_{ij}=\ell^{'}_{ij}/\ell_{ij}$, called the length factor, 
for each node pair $ij$~\cite{Blunden}, which measures the fractional increase of the path between 
$i$ and $j$, and explore the distribution $\Phi(\alpha)$ 
and its cumulative $F(\alpha)=\int^{\alpha}\Phi(\alpha^{'})d\alpha^{'}$
over realizations of $P_a(w)$ (and $\mathcal{G}$ for ER networks).
Note that the LPP reachability condition is $\alpha_{ij}\leq\tau$.

\begin{figure}
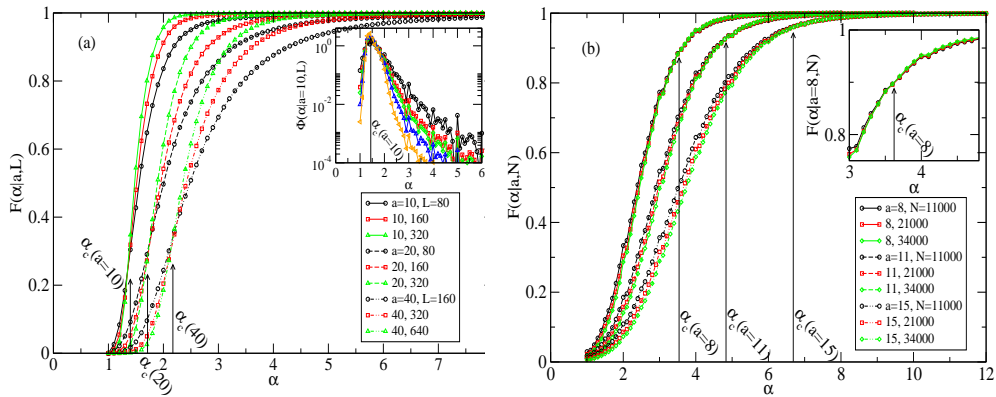

\epsfig{file=fig2a.eps,width=6.5cm,height=5.2cm}\vspace{1cm}
\epsfig{file=fig2b.eps,width=6.5cm,height=5.2cm}
\caption{Cumulative $F(\alpha|a,L)$ vs. $\alpha$ for 2-dimensional lattices, and $F(\alpha|a,N)$ for ER networks. 
Plot (a) for lattices corresponds to different $L$ and $a$.
The inset shows $\Phi(\alpha|a,L)$ with logarithmic scale in the
vertical axis, emphasizing the fast decay of the distribution which is even more pronounced for larger system sizes.
This supports the idea that, at the asymptotic limit $L\rightarrow\infty$, 
the distribution is highly concentrated around $\alpha_c$ (highlighted by the arrow). Note that the location
of $\alpha_c$ also corresponds to the location where $F(\alpha|a,L)$ cross over each other as $L$ increases.
For increasing $a$, the location of $\alpha_c$ shifts to the right.
Panel (b) for ER networks corresponds to $F(\alpha|a,N)$ for several $N$ and $a$ (distribution $\Phi$ is omitted due 
to discretization fluctuations). $\alpha_c$ is such that $F(\alpha|a,N)$ is approaching 1.
Increasing $a$ leads to distributions shifted to the right. The inset shows the crossing over between
distributions with fixed $a$ and increasing $N$.
}
\label{Phialpha-latt}
\end{figure}

Figure~\ref{Phialpha-latt}(a) shows the measurement of $F(\alpha|a,L)$ in lattices, and in the inset
the distribution $\Phi(\alpha|a,L)$, which is a well
concentrated function around its maximum $\alpha=\alpha_c$.
The cumulative $F(\alpha)$ increases sharply around $\alpha=\alpha_c$ rapidly 
approaching 1, which indicates that many node pairs satisfy $\alpha\leq\alpha_c$.
The sharpness of $\Phi(\alpha)$ increases for larger $L$, and $F(\alpha)$ becomes more step like, 
while $\alpha_c$ remains in the same location. Note that $\alpha_c$ 
can also be determined at the location where $F(\alpha|a,L)$ for increasing $L$ cross over each other.
Increasing $a$, on the other hand, leads to increasing $\alpha_c$, consistent
with path lengths becoming longer under more disordered conditions.
In Fig.~\ref{Phialpha-latt}(c), we focus on $F(\alpha|a,N)$ for ER networks, and observe similar
features to those in the case of lattices (we display the cumulative only as
the small values of path lengths in ER graphs produce large discretization fluctuations),
apart from the asymptotics which appear to be slower.
\subsection{Scaling of path length as a function of disorder}
In order to understand the previous results, we consider the
current knowledge on the problem of optimal paths, which has received
considerable attention in the context of surface growth and domain 
walls~\cite{Chen,Huse-Henley,Cieplak} in the physics literature. For the
purpose of clarity we briefly review these results here, starting with
lattices and extending the discussion to networks. 

Disordered lattices often exhibit
optimal paths which are self-affine, characterized by
lengths which scale linearly with $\ell_{ij}$,
with a constant prefactor dependent on the roughness exponent related to
the disorder~\cite{Huse-Henley}.
This limit of self-affine paths has become known as the weak
disorder limit.

Another scaling regime has been recognized~\cite{Cieplak} when
the disorder approaches the so-called strong disorder limit.
In this limit, each link weight in the network is very different to any other link weight,
progressively forcing the optimal paths to lie inside the minimum
spanning tree where their lengths scale as $\ell^{d_{\rm opt}}_{ij}$,
$d_{\rm opt}$ being the scaling exponent of the
shortest path in the minimum spanning tree~\cite{Chen,Porto,Ioselevich}.

A general theory explaining these optimal path limits
points out that weak and strong disorder are separated by 
a disorder length-scale $\xi$ which depends on the disorder distribution 
$P_a(w)$ and some lattice dependent features~\cite{Chen,Wu}.
Optimal paths covering a distance 
smaller than $\xi$ are in strong disorder, and those covering a larger distance are in weak disorder
(provided the system is large enough so that $\xi\ll L$). 
Thus, $\xi$ is the weak-strong disorder crossover length. The weak and strong disorder
scaling regimes for $\ell^{'}_{ij}$ can be expressed by the scaling relation
\begin{equation}
\ell^{'}_{ij}\sim\xi^{d_{\rm opt}} f\left(\frac{\ell_{ij}}{\xi}\right),
\quad
f(x)=
\left\{
\begin{array}{ll}
x,&x\gg 1(\text{weak})\\
\text{const.},&x\ll 1 (\text{strong}).
\end{array}
\right.  \label{lopt}
\end{equation}
The length-scale $\xi$ can be determined by the relation
$\xi=[p_c/(w_c P_a(w_c))]^{\nu}$~\cite{Chen}, where $p_c$ is the percolation
threshold of the lattice, $\nu$ the correlation length exponent of percolation, and 
$w_c$ is the solution to the equation $p_c=\int^{w_c}P_a(w)dw$, i.e., the weight
for which the cumulative distribution of $P_a(w)$ is equal to $p_c$. 

\begin{figure}
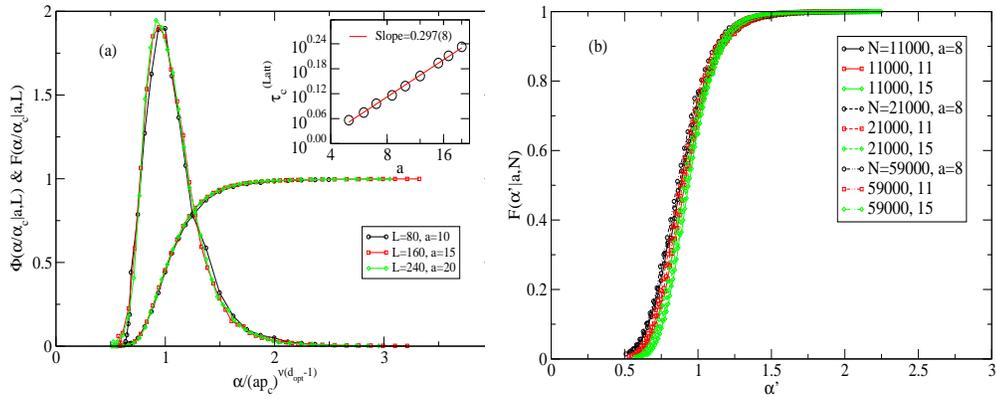

\epsfig{file=fig3a.eps,width=6.5cm,height=5.2cm}\vspace{1.5cm}
\epsfig{file=fig3b.eps,width=6.5cm,height=5.2cm}
\caption{Testing Eqs.~(\ref{tau-alpha_c}) and (\ref{beta}).
(a) $\Phi(\alpha/\alpha_c)$ and $F(\alpha/\alpha_c)$ vs $\alpha/\alpha_c$ for lattices with 
$\alpha_c\sim (ap_c)^{\nu(d_{\rm opt}-1)}$, for various $a$ and $L$ specified in the legend.
Both $\Phi$ and $F$ collapse supporting the hypothesis that $\alpha_c$ responds to Eq.~(\ref{alphac-latt})
for lattices. Inset: $\tau_c^{\rm(Latt)}$ vs. $a$ from data and least squares fit. The relation predicted by our
theory is $\tau_c\sim (ap_c)^{\nu(d_{\rm opt}-1)}$ which is equal to
0.293 from independent work, within error of $0.297\pm 0.008$ obtained from the fit from our simulation results.
(b) $F(\alpha^{'}|a,N)$ vs. $\alpha^{'}$ for ER networks for a combination of $N$ and $a$. The collapse of the
curves strongly supports Eq.~(\ref{beta}).
}
\label{Phialpha-scaled}
\end{figure}

Based on the previous arguments, we can now postulate the properties of
$\Phi(\alpha)$. We concentrate on the weak disorder limit because it is the only possible
regime in which an LPP transition could take place~\cite{whynotstrong}. In this
regime, based on Eq.~(\ref{lopt}), we find that 
$\ell^{'}_{ij}\sim \xi^{d_{\rm opt}}(\ell_{ij}/\xi)=\xi^{d_{\rm opt}-1}\ell_{ij}$, 
where the parenthesis corresponds to $\ell_{ij}$ in the scale of the crossover length 
$\xi$, and $\xi^{d_{\rm opt}}$ to the length of the path within this crossover length
(see Fig.~3 in Ref.~\cite{Wu}).
This relation indicates that the typical $\alpha$ 
is given by $\alpha_c\sim\ell^{'}_{ij}/\ell_{ij}=\xi^{d_{\rm opt}-1}\ell_{ij}/\ell_{ij}=\xi^{d_{\rm opt}-1}$. 
For $P_a(w)$ of Eq.~(\ref{Pw}), $\xi=(ap_c)^{\nu}$ calculated according to the above formalism, producing
\begin{equation}
\alpha_c^{(\rm Latt)}\sim (ap_c)^{\nu(d_{\rm opt}-1)}.
\label{alphac-latt}
\end{equation}
In order to test this, we present in Fig.~\ref{Phialpha-scaled}(a) (main)
the scaled curves $\Phi(\alpha/\alpha_c)$ for various $L$ and $a$,
where $\alpha_c$ is taken from Eq.~(\ref{alphac-latt}). The collapse is
consistent with our scaling picture, and indicates that indeed there is
a clear path length increase $(ap_c)^{\nu(d_{\rm opt}-1)}$ that explains 
the empirical results.

Given the large fraction of node pairs for which $\alpha$ is close to $\alpha_c$,
we postulate that 
\begin{equation}
\tau_c^{(\rm Latt)}=\alpha_c^{(\rm Latt)}(a)\sim (ap_c)^{\nu(d_{\rm opt}-1)},
\label{tau-alpha_c}
\end{equation}
i.e., the tolerance necessary to obtain the LPP transition is equivalent to the most probable length factor.
To test this relation, we find by inspection the values of $\tau_c^{\rm(Latt)}$ as a function of $a$, and plot
them in Fig.~\ref{Phialpha-scaled}(a) (inset). The relation between $\tau_c^{\rm(Latt)}$ and $a$ can be fit to 
a power law with exponent
$0.297\pm 8$, which is very close to the predicted $\nu(d_{\rm opt}-1)=0.293$, strongly supporting 
Eq.~(\ref{tau-alpha_c}). In addition, we find that $\sigma^*(\tau_c^{(\rm Latt)})\sim 0.2\pm0.04$, independent
of $a$.
The previous results indicate that $\sigma^*$ is universal. We have tested this by comparing the shape of
$\sigma^*(\tau,a,L)$ for different $a$, $\tau$, and $L$, and have found that adjusting for a given
combination of these values, the curves for $\sigma^*$ can be made to overlap, supporting universality.

For ER networks, it is known that their path length distributions
are concentrated due to their random structure, leading to a large number of lengths being similar 
to an overall typical length~\cite{Braunstein}.
Thus, we simplify our analysis by focusing on the typical lengths
before and after the introduction of disorder. Before disorder sets in, the typical path
length is
\begin{equation}
\ell_{ER}\sim\frac{\log N}{\log\langle k\rangle}.
\end{equation}
However, after weak disorder sets in, it becomes
\begin{equation}
\ell^{'}_{ER}\sim \mu ap_c \log\left(\frac{N^{1/3}}{ap_c}\right)
\end{equation}
which emerges from the relation that is the equivalent to Eq.~(\ref{lopt})
now applied to networks. $\mu$ is a quantity not yet characterized in the literature, and depends on $\langle k\rangle$ 
and also weakly on $N$ and $a$; for our simulations, $\mu\approx 4$. 
The $N$-dependent ratio between $\ell_{ER}$ and $\ell^{'}_{ER}$, which we label $\beta(N)$ (slightly different
to $\alpha$ because the later applies to each pair of nodes, but the former to the overall typical distances),
is given by
\begin{equation}
\beta(N)\sim \mu ap_c\log\langle k\rangle \left[\frac{1}{3}-\frac{\log(ap_c)}{\log N}\right]=
\alpha_{c,\infty}^{\rm(ER)}-\epsilon^{\rm(ER)}(N)
\label{beta}
\end{equation}
where $\alpha_{c,\infty}^{\rm(ER)}\equiv\mu ap_c\log\langle k\rangle/3$, the asymptotic value of $\beta$, and
$\epsilon^{\rm(ER)}(N)\equiv\mu ap_c\log\langle k\rangle\log(ap_c)/\log N$, a finite size correction that vanishes
logarithmically with $N$ as $N\rightarrow\infty$.
In analogy to lattices, we hypothesize that $\tau_c^{\rm(ER)}(N)$ scales as $\beta(N)$, which includes the size 
corrections with $N$. Also, in the limit
$N\rightarrow\infty$, we define $\tau_{c,\infty}^{\rm(ER)}\equiv\alpha_{c,\infty}^{\rm(ER)}$. 

To test Eq.~(\ref{beta}), we introduce the rescaled variable $\alpha^{'}_{ij}=(\alpha_{ij}+
\epsilon^{\rm(ER)}(N))/\alpha_{c,\infty}^{\rm(ER)}$, and plot
$F(\alpha^{'}|a,N)$ in Fig.~\ref{Phialpha-scaled}(b).
The collapse of the curves is excellent, and supports the validity of our assumptions.
In the thermodynamic limit we expect
\begin{equation}
\tau_{c,\infty}^{(ER)}=\alpha_{c,\infty}^{(ER)}\sim \frac{\mu ap_c\log\langle k\rangle}{3},
\end{equation}
but it is important to keep in mind the large
finite size corrections to $\tau_c^{\rm(ER)}$, which make it more similar in value to Eq.~(\ref{beta})
for finite $N$.
\section{Conclusions}
We study disorder induced Limited Path Percolation
and determine that a percolation transition occurs for a critical tolerance $\tau_c$.
The critical tolerance increases together with the heterogeneity of the disorder.
Numerical results indicate that LPP displays universality. Also, the phase transition is first order, 
with the discontinuity in the order parameter $\sigma^*(\tau_c)$ independent of the disorder.
For lattices $\sigma^*(\tau_c)$ is of the order of 0.2; for Erd\H{o}s-R\'enyi networks, our numerical 
results suggest it may approach 1 signaling a catastrophic transition.
The concept of length factor applied to the theory of optimal paths predicts
a typical factor $\alpha_c$ which, in turn, predicts $\tau_c$.
We believe the tolerance thresholds predicted here reflect reachability conditions
of some real networks under real failure scenarios (congestion, maintenance, etc.)
better than regular percolation.
\acknowledgments
E.L. was supported by TSB grant SATURN (TS/H001832/1). L.A.B. was supported by UNMdP and FONCyT (PICT/0293/08).
E.L. thanks Felix Reed-Tsochas and Austin Gerig for helpful discussions.

\clearpage

\clearpage

\begin{thebibliography}{99}
\bibitem{no-time-depen}
The details of how these 
weights emerge and their typical range of values depend on the specific problem, and we will not
address this problem here, because it is fundamentally context dependent.
\bibitem{fnpgtrpc} For LPP to occur, $p>p_c$, as there must at least be a 
connected cluster when no length condition is imposed before reachable
clusters can be found under the more strenuous condition of limited path lengths.
\bibitem{reg-LPP} Regular LPP is equivalent to choosing $1-p$ links to have $w=\infty$, while the others remain at $w=1$.
\bibitem{whynotstrong} The strong disorder limit cannot produce a disorder induced LPP transition
because the optimal paths scale as a power law of the original path. Any length condition
that relies on the disordered and ordered paths being linear with each other is only found in the weak disorder
limit.
\end{thebibliography}
\end{document}